\documentclass[prb,showpacs,floatfix,twocolumn]{revtex4}

\usepackage{graphicx}

\newcommand{\zmo}{ZnMn$_2$O$_4$}
\newcommand{\zmox}{Zn$_x$Mn$_{3-x}$O$_4$}
\newcommand{\zmotwo}{Zn$_{0.25}$Mn$_{2.75}$O$_4$}
\newcommand{\zmofiv}{Zn$_{0.5}$Mn$_{2.5}$O$_4$}
\newcommand{\zmosev}{Zn$_{0.75}$Mn$_{2.25}$O$_4$}
\newcommand{\haus}{Mn$_3$O$_4$}

\begin{document}

\title{Intrinsic exchange bias in \zmox\, ($x \leq 1$) solid solutions}

\author{Daniel P. Shoemaker} \email{dshoe@mrl.ucsb.edu}
\author{Efrain E. Rodriguez}
\author{Ram Seshadri}
\affiliation{Materials Department and Materials Research Laboratory\\
University of California, Santa Barbara, CA, 93106, USA}
\author{Ivana Sabaj Abumohor} 
\affiliation{Centro para la Investigaci\'{o}n Interdisciplinaria Avanzada en 
Ciencia de los Materiales, Departamento de Ingenier\'{i}a Qu\'{i}mica y 
Biotecnolog\'{i}a, Universidad de Chile, Casilla 2777, Santiago, Chile} 
\author{Thomas Proffen}
\affiliation{Los Alamos National Laboratory, Lujan Neutron Scattering Center, MS H805, Los Alamos, New Mexico 87545, USA}

\date{\today}

\begin{abstract}

Bulk specimens of the het\ae rolite solid solution \zmox\, with $x$ = 0, 0.25, 
0.5, 0.75, and 1 have been prepared as homogeneous, phase-pure polycrystalline
samples as ascertained by neutron diffraction measurements. Samples with $x$ = 
0.25, 0.5, and 0.75 exhibit shifted magnetic hysteresis loops at low 
temperature, characteristic of exchange bias typically seen in magnetic 
composites. We propose that the unusual magnetic behavior arises as a result 
of a nanoscale mixture of ferrimagnetic and antiferromagnetic regions that are 
distinct but lack long-range order. While some glassy behavior is seen 
in AC magnetic measurements, its magnitude is not sufficient to account for 
the observed dramatic exchange bias. Furthermore, isothermal and thermoremanent 
magnetization measurements distinguish this material from a pure spin 
glass. The title system offers insights into the alloying of a ferrimagnet 
\haus\, with an antiferromagnet \zmo\, wherein distinct magnetic clusters grow 
and percolate to produce a smooth transition between competing orders.

\end{abstract}

\maketitle 

\section{Introduction} 

Exchange bias is a magnetic memory effect that occurs at the interface between 
a ferromagnet (or ferrimagnet) and an 
antiferromagnet.\cite{meiklejohn_new_1957} By field-cooling a system with an 
ordered ferromagnet/antiferromagnet interface through the N\'{e}el temperature 
$T_N$ of the antiferromagnet, exchange interactions at the interface lead to a 
preferred direction of magnetization, typically along the cooling field 
direction. Exchange bias has been engineered into a wide variety of materials 
systems and geometries: core-shell nanoparticles, granular composites, and thin
film read-heads for magnetic recording media.\,\cite{nogues_exchange_2005}
In addition to the abrupt interfaces in thin-film architectures, a significant 
thrust has been made toward understanding the mechanisms of loop-shifting 
phenomena in disordered and composite magnets.
 
Disordered and/or dilute magnetic spins in a crystal can lead to glassy 
behavior that gives rise to magnetic memory effects as a result of 
slow and time-dependent processes below the spin freezing temperature $T_f$.
Such glassiness can result in biased magnetization loops. Distinctions 
between exchange bias and glassy magnetism are therefore useful.  
Exchange-biased systems are usually expected to have (i) two magnetic phases 
with a well-defined interface, (ii) a loop shift, measured as the exchange 
field, $H_E$, that goes to zero above $T_N$, and (iii) zero exchange field (loop 
shift) if the cooling field is zero; exchange bias is not observed for 
$M-H$ loops acquired after zero-field cooling. Spin glasses, in turn, are 
associated with (i) frozen spins below $T_f$ that produce a frequency-dependent
peak in susceptibility, (ii) an absence of long-range magnetic ordering, and 
(iii) some relaxation on a macroscopic time scale after field changes below 
$T_f$.\cite{fischer_spin_1985,binder_spin_1986}
 
As an illustrative example, loop shifts along the field axis were observed in 
the prototypical spin glass CuMn by Monod, \textit{et al.} in 
1979,\cite{monod_magnetic_1979} but these are not strictly considered to be 
evidence for exchange bias since the magnetic phase is homogeneous and 
field-cooling is not necessary. A glassy phase can occasionally fulfill the 
role of an antiferromagnet in a two-phase exchange biased system: loop shifts are 
commonly observed in ferromagnetic-core nanoparticles with disordered 
surface layers, where a spin-glass-like relaxation of the remanent 
magnetization versus time is accompanied by a loop 
shift.\,\cite{kodama_surface_1996,makhlouf_magnetic_1997,martinez_low_1998} 
Glassy spins freeze to partially align with the ferromagnetic spins during 
field cooling and a preferred direction of magnetic orientation is therefore 
imparted. A detailed study of the interplay between 
ferromagnet/spin glass Co/CuMn bilayers with well-defined thicknesses has confirmed this 
behavior.\,\cite{ali_exchange_2007}

Here we report a detailed study of the magnetic properties of \zmox\, 
($x \leq 1$) solid solutions, studied in phase-pure polycrystalline samples.
This system was reported many decades ago by Jacobs and 
Kouvel,\cite{jacobs_exchange_1961} who found that exchange bias and ``magnetic 
viscosity'' effects (meaning glassy magnetism in the current context) were 
found to occur together in the solid solution. We
re-examine this system in light of the increased interest in nanoscale 
inhomogeneities in functional, and particularly correlated 
oxides.\cite{dagotto,mathur} We focus in particular on the role of magnetic 
inhomogeneities and how they result in competing magnetic order. We probe
the question of whether these magnetic inhomogeneities are assocated with 
structural inhomogeneities, in the sense of the formation of nanocomposite
architectures. We also examine the nature of glassy magnetism in this system 
and make distinctions between glassiness and exchange bias.
 
The end members hausmannite Mn$_3$O$_4$ and het\ae rolite \zmo\,are a spiral
ferrimagnet and an antiferromagnet, respectively, with the former compound 
having recently emerged as a candidate magnetoelectric material as a 
consequence of its complex magnetic 
ordering.\cite{tackett_magnetodielectric_2007} At high temperatures 
($>$1100$^\circ$C) these compounds are cubic spinels, but they distort to the 
tetragonal het\ae rolite structure below 1100$^{\circ}$C as a consequence of 
orbital ordering of octahedral $d^4$ Mn$^{3+}$, as first described by 
Goodenough.\cite{goodenough_spinels_1955,goodenough_manganites_1955}
The octahedral site is completely occupied by Mn$^{3+}$. The tetrahedral site 
accommodates alloying of isovalent $d^{10}$ Zn$^{2+}$ and $d^5$ Mn$^{2+}$,
the former being an ion that prefers tetrahedral coordination, and the latter,
an ion that lacks a site preference.  

We find, in agreement with, but significantly extending the original work of 
Jacobs and Kouvel, \cite{jacobs_exchange_1961} that \zmox\, does not behave 
like a random solid solution in the magnetic sense, and neither does it 
macroscopically phase-separate into \zmo\,and \haus. Instead, features of both 
are present, and the complex magnetic behavior can be explained by invoking
nanoscale clusters of ferrimagnetic spins that gradually grow and percolate 
as $x$ is increased. These nanoscale ferrimagnetic regions always abut 
nanoscale antiferromagnets for $x < 1$ and this results in the observed 
exchange bias.
 
Intrinsic exchange biased systems have similarly been reported in perovskite 
manganites and cobaltites with mixed valent 
B-sites.\cite{tang_glassy_2006,luo_response_2008}
For example, the system (Y,Sr)MnO$_3$ has been reported as displaying 
glassiness as well as loop shifting.\,\cite{karmakar_evidence_2008}
In contrast to these perovskite systems, we find striking magnetic complexity 
in the title solid solution in the absence of any site disorder on the B-site.
Additionally, the solid solution does not require aliovalent substitution and
concomitant changes in the valence states of ions.

In general, the magnetic structure of spinel compounds such as \zmo\, can be 
influenced in two ways: through tuning the average size of cations in the 
tetrahedral site, and through the addition of spins on the tetrahedral A site. 
Such tuning \textit{via} the A-site cation radius has been studied extensively
in chalcogenide spinels, but rarely changes the type of magnetic ordering in
oxide spinels.\,\cite{rudolf_spin-phonon_2007, martinho_magnetic_2001} 
Tuning \textit{via} the introduction of magnetism on the A-site has been
studied in the (Zn,Co)Cr$_2$O$_4$ system.\cite{melot_magnetic_2009}
In these Cr oxide spinels, like in the title Mn spinels, the B site is always 
occupied by Cr$^{3+}$ or Mn$^{3+}$. In cases where B-site Mn$^{3+}$ is alloyed 
with non-Jahn-Teller ions, dramatic phase separation due to dilution of the 
orbital ordering patterns is observed.\cite{yeo_solid_2006,yeo_generic_2009}

\section{Methods}

Ceramic pellets of \zmox\,were prepared by grinding stoichiometric amounts of 
ZnO and MnO ( both 99.9\,\% from Aldrich) in an agate mortar and pestle, 
pressing at 100\,MPa, and firing in air at temperatures between 950$^{\circ}$C 
and 1200$^{\circ}$ for 24\,h (water quenched for $x$ = 0 and 0.25) in 
accordance with the phase diagram of Driessens and 
Rieck.\cite{driessens_phase_1966} For all calcinations, pellets were buried in 
sacrificial powder of the same composition in covered alumina crucibles. 
The purity of all samples was confirmed by laboratory X-ray diffraction (XRD) 
data acquired on a Philips X'Pert diffractometer with Cu-$K_\alpha$ radiation. 
Magnetic properties were measured using a Quantum Design MPMS 5XL SQUID 
magnetometer. Time-of-flight (TOF) neutron powder diffraction on samples 
held in vanadium cans at the high intensity powder diffractometer (HIPD) at 
Los Alamos National Laboratory. The HIPD instrument can collect high 
$d$-spacing magnetic reflections out to tens of \AA. However, no peaks were
found beyond 6\,\AA\, in any of the samples studied here. We limit the 
Rietveld refinement to banks 1--4, with a maximum momentum transfer 
$Q_{max}$ = 20\,\AA$^{-1}$ and maximum $d$-spacing of 6\,\AA.
Rietveld refinement was performed using the XND code\cite{berar_xnd_1998} for 
X-ray data and GSAS\cite{larson_general_2000} for TOF data. Crystal structures 
are visualized using VESTA.\cite{momma_vesta_2008}

\section{Results and Discussion}

\begin{figure}
\centering\includegraphics[width=8cm]{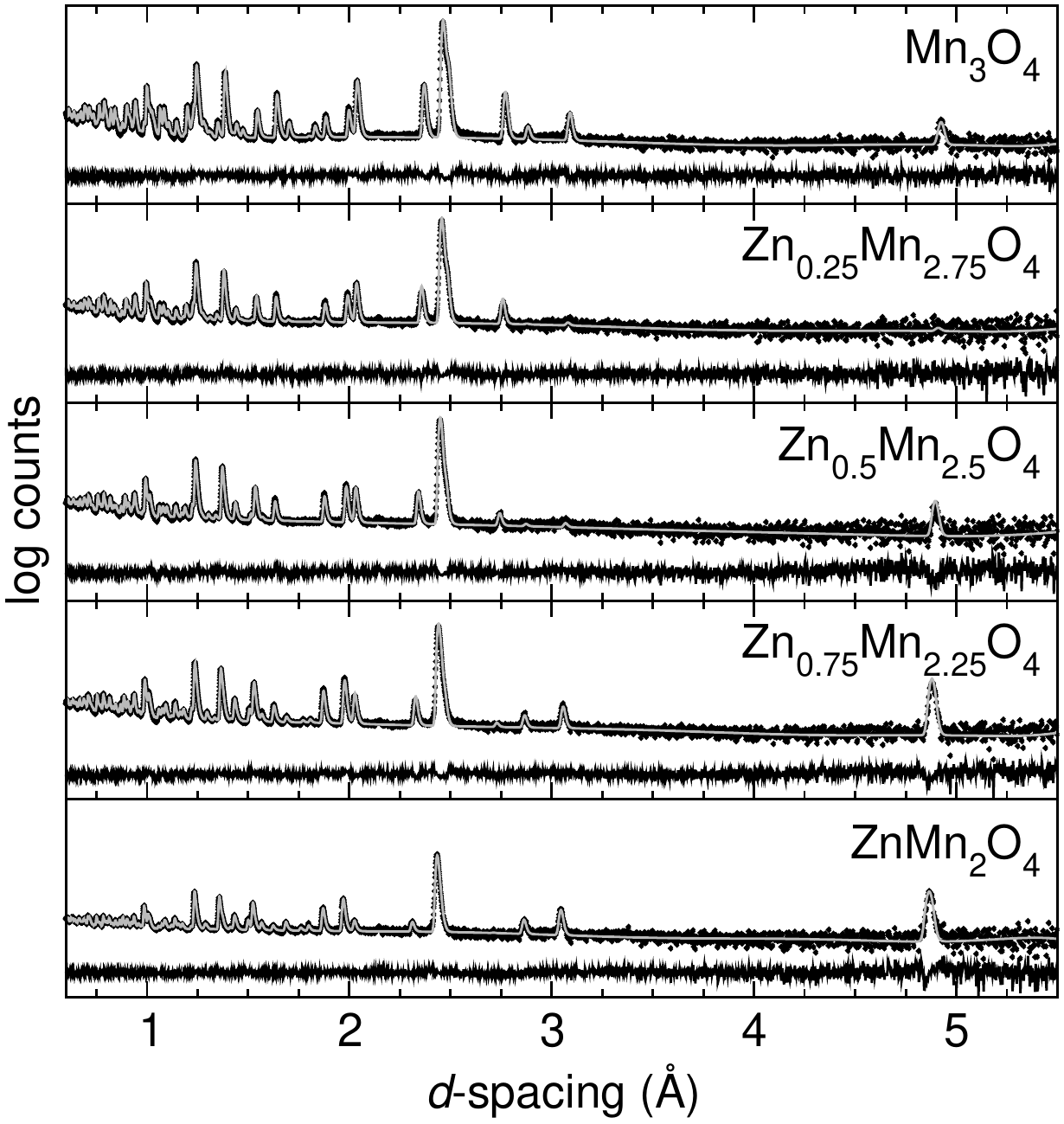}\\
\caption{300\,K neutron TOF diffraction Rietveld refinements in the 
$I4_1/amd$ space group confirm the purity of all \zmox\,phases at 300\,K. 
Difference profiles are shown below each fit. Refinement results (including 
$R_{wp}$) are provided in Table \ref{tab:rietveld}.}
\label{fig:xrd}
\end{figure}

\begin{figure}
\centering\includegraphics[width=8cm]{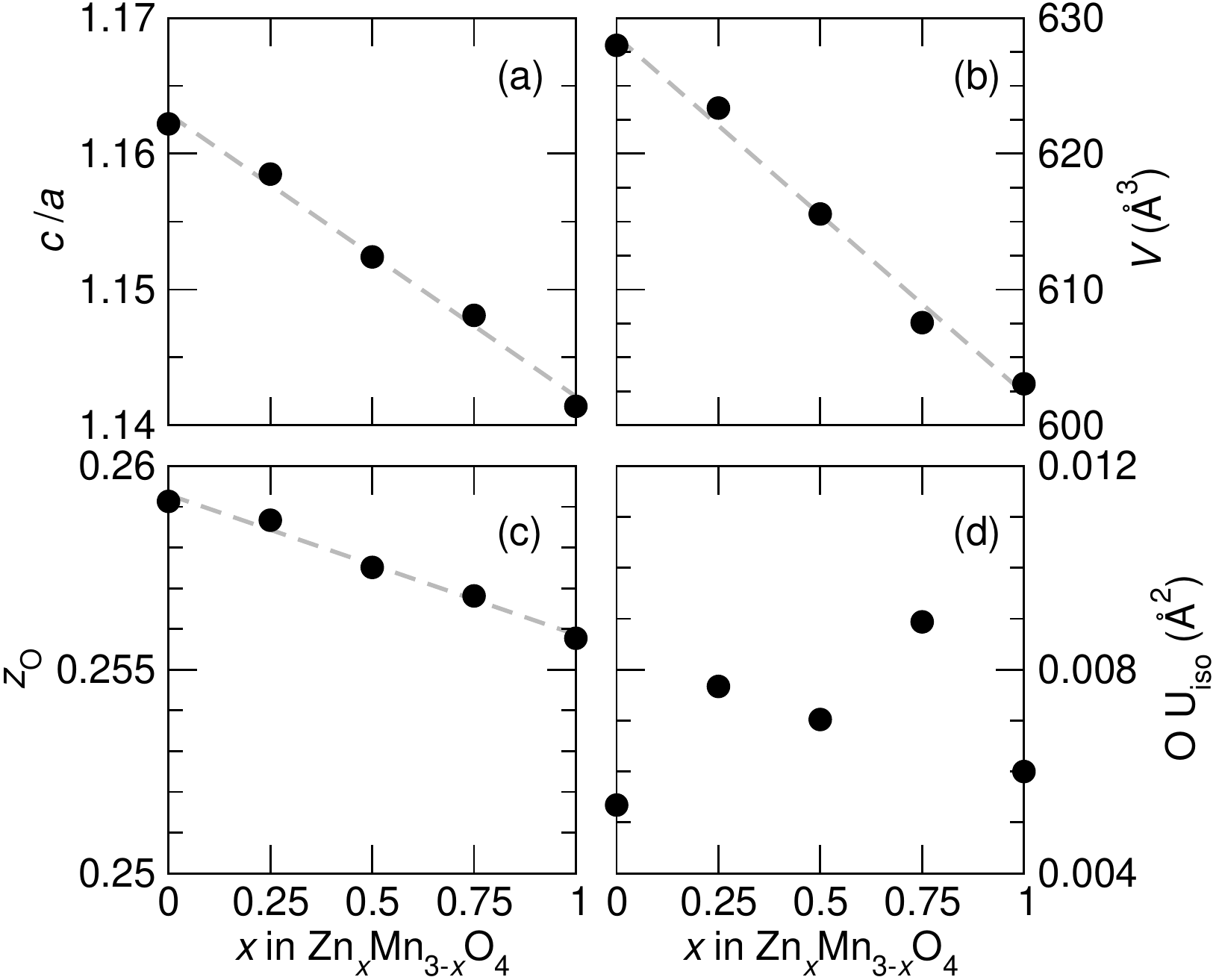}\\
\caption{Structural parameters at 300\,K from neutron TOF Rietveld refinements show 
decreasing (a) $c/a$ ratios and (b) cell volume with Zn concentration (linear 
fits, dashed), due to its smaller radius. The oxygen $z$ position in (c) 
decreases toward the undistorted value of 0.25. In (d), chemical disorder 
causes compounds with intermediate Zn/Mn mixing to have higher thermal 
parameters than the end members. Error bars are smaller than the symbols in 
all panels.} 
\label{fig:latparams}
\end{figure}

Time-of-flight neutron diffraction is an especially useful tool in examining 
the solid solutions studied here. In addition to the possibility of variable 
temperature studies, the availability of high resolution high momentum transfer
($Q$) data, the ability to probe magnetic scattering, and the ability to 
examine Zn$^{2+}$/Mn$^{2+}$ A-site distribution are all advantageous.
The nuclear scattering lengths are 5.68\,fm for Zn and $-$3.73\,fm for Mn,
so these ions are extremely well contrasted in the scattering.

\begin{figure}
\centering\includegraphics[width=9cm]{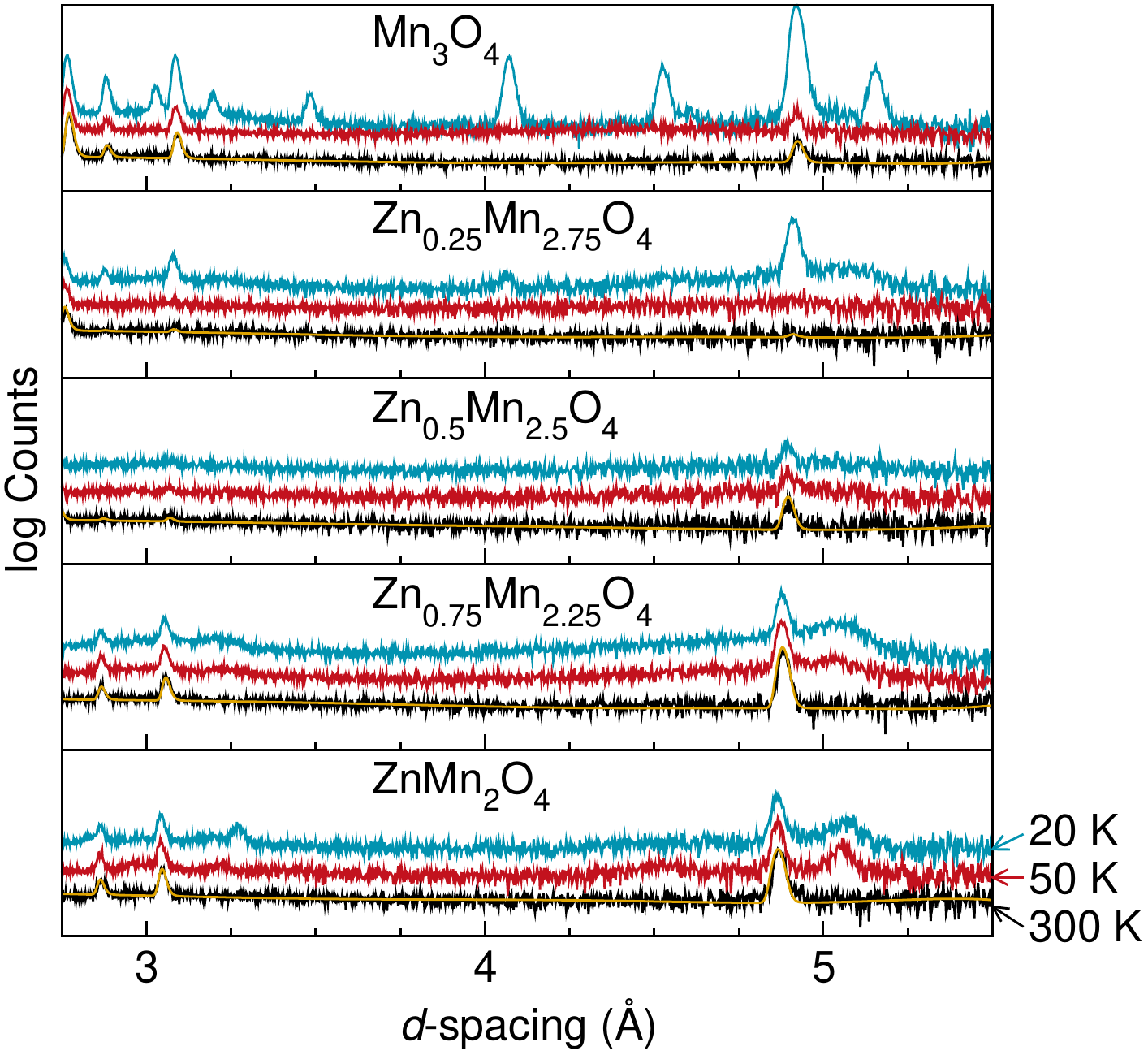}\\
\caption{(Color online) Neutron TOF powder diffraction patterns 
(log scale, offset for clarity) for the \zmox\, solid solutions at 300\,K, 
50\,K and 20\,K. The Rietveld fit to the 300\,K (non-magnetic) profile is 
shown for all samples. Note that only diffuse magnetic scattering is evident 
around $d$ = 5\,\AA\, for the sample with $x = 0.5$. In \haus, the baseline at 
20\,K drops due to transfer of diffuse magnetic scattering to Bragg peaks.} 
\label{fig:hipd}
\end{figure}

Room temperature neutron TOF diffraction patterns are shown in 
Fig.\,\ref{fig:hipd}, along with fits to the profiles using the Rietveld 
refinement method. The fits give excellent matches to the het\ae rolite 
structure across the solid solution. The TOF refinements reveal no impurities,
and the particles are many microns in extent as seen from the narrow widths of 
the diffraction peaks. Structural parameters from the Rietveld refinement are 
provided in Table\,\ref{tab:rietveld}. Trends in the relevant structural 
parameters as a function of $x$ are shown in Fig.\,\ref{fig:latparams}. 

The cell volume and $c/a$ ratios vary smoothly and reflect the 10\,\% 
difference in the ionic radii of tetrahedral Mn$^{2+}$ (0.66\,\AA) and 
tetrahedral Zn$^{2+}$ (0.60\,\AA). The decrease in tetragonality as the Zn 
content $x$ increases could be due to its preference for covalent bonding, and 
therefore a tendency toward more regular tetrahedral coordination.
This is supported by the oxygen $y$ and $z$ coordinates, which approach their 
least-offset values of $\frac14$ and $\frac12$, respectively with increasing Zn.
The oxygen $U_{\textrm{iso}}$ values for each compound are relatively close, 
but the smallest values occur for the end members, while site mixing on the 
A-site leads to larger values for intermediate $x$. Random A-site mixing of 
Zn$^{2+}$/Mn$^{2+}$ is suggested by the smoothly varying lattice parameters 
and the $c/a$ ratios versus $x$. This system strictly maintains a ``normal'' 
distribution of cations: Zn$^{2+}$ greatly prefers tetrahedral coordination by 
oxygen, and Mn$^{3+}$ is very stable in a JT distorted octahedral 
coordination.\cite{miller_distribution_1959} The A-site occupation refines to 
within 1\,\% of the nominal Zn/Mn ratio in each case. The JT distortion is 
present in all samples since the B sublattice is invariant with composition 
$x$.\,\cite{goodenough_theory_1955}

\begin{table*}
\caption{\label{tab:rietveld} 
Bulk structural parameters at 300\,K for \zmox\,obtained from Rietveld refinement of 
TOF neutron diffraction data: $I4_1/amd$ (No. 141, origin choice 2); 
A-site Zn$_x$Mn$_{1-x}$ at (0,$\frac14$,$\frac78$); B-site Mn at 
(0,$\frac12$,$\frac12$); O at (0,$y$,$z$).}
\centering
\begin{tabular}{lllllllll}
\hline
Composition&$a$ (\AA)&$c$ (\AA)&$c/a$&$y_O$&$z_O$&O $U_{\textrm{iso}}$ (\AA$^2$) & $R_{wp}$ (\%)	\\
\hline
\hline
\zmo    &5.71643(5)&9.2275(1) &1.1414&0.47657(8)&0.25577(5)&0.0060(2) &3.1 \\
\zmosev &5.71955(3)&9.28628(7)&1.1481&0.47524(3)&0.25681(2)&0.00894(4)&2.8 \\
\zmofiv	&5.73726(3)&9.3504(1) &1.1524&0.47499(3)&0.25751(2)&0.00702(4)&3.0 \\
\zmotwo &5.75134(4)&9.4225(1) &1.1585&0.47404(4)&0.25867(3)&0.00767(4)&3.3 \\
\haus   &5.75924(2)&9.46632(6)&1.1622&0.47273(3)&0.25913(2)&0.00534(7)&2.7 \\
\hline
\hline
\end{tabular}
\end{table*}

Figure \ref{fig:hipd} displays TOF diffraction patterns at $T$ = 300\,K, 50\,K,
and 20\,K over a region that contains all magnetic scattering intensity 
relevant to the discussion here. Most obvious are the numerous, intense 
magnetic peaks in the end member Mn$_3$O$_4$. The top panel is on a 
log scale one order of magnitude higher than the rest. The onset of long-range 
magnetic ordering leads to a transfer of intensity from the diffuse scattering 
to the Bragg peaks, resulting in a much lower baseline for the 20\,K data than 
that at higher temperatures.\,\cite{kuriki_diffuse_2003} The magnetic structure
of hausmannite Mn$_3$O$_4$ is complex, with the onset of incommensurate 
sinusoidal magnetic ordering at $T_C$ = 44\,K, followed by a locking in of the 
spin modulation to a commensurate structure 
below 33\,K.\,\cite{jensen_magnetic_1974, chardon_mn3o4_1986}

\begin{figure}
\centering\includegraphics[width=8cm]{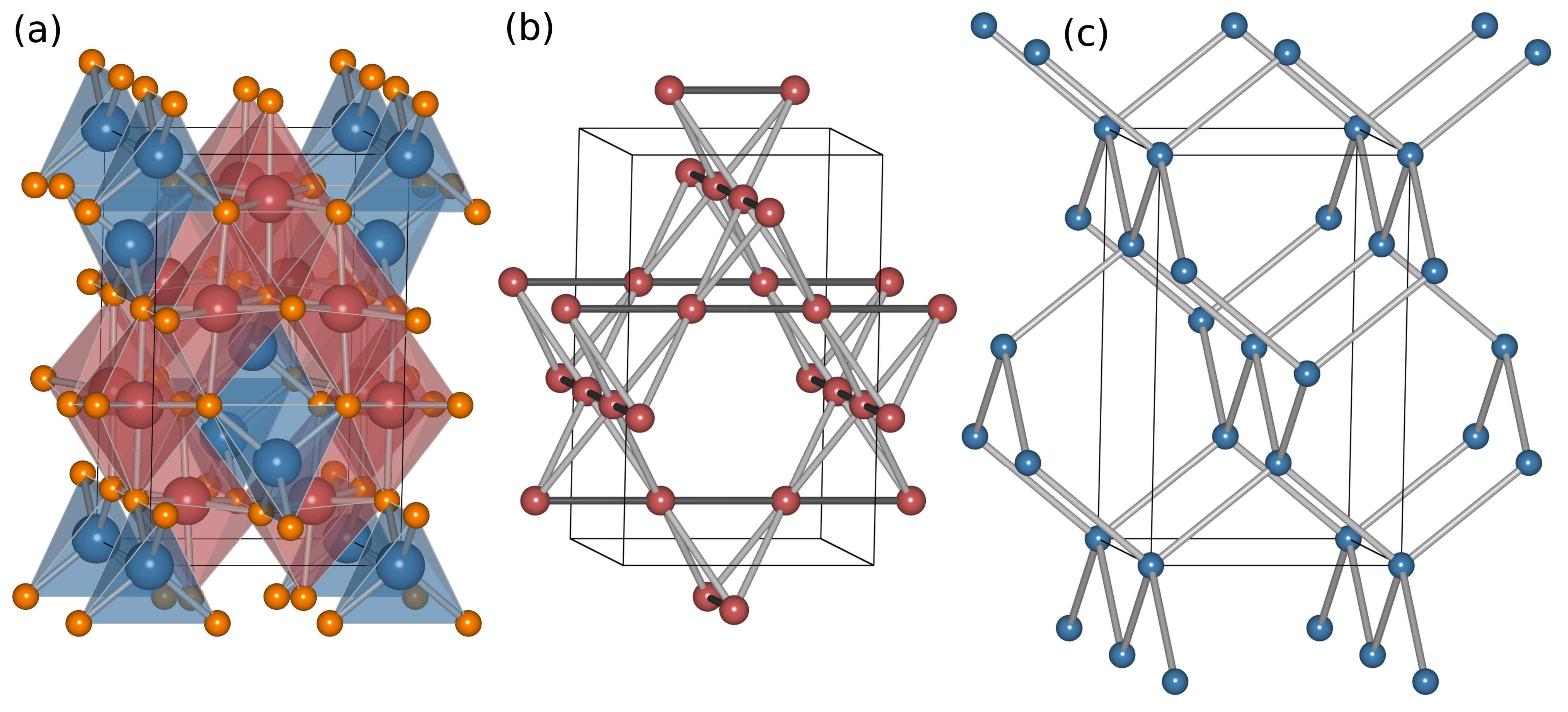}\\
\caption{(Color online) The \zmo\,het\ae rolite unit cell is shown in (a)
with oxygen polyhedra drawn around Mn (red) and Zn (blue). In (b), the B-site
linkages are shown. The B--B direct exchange net consists of a stretched
pyrochlore lattice (four interwoven kagom\'{e} nets) with B--B links in $a$
and $b$ directions (dark) that are shorter than those with a $c$ component
(light). The diamond-type A lattice is shown in (c).} 
\label{fig:structure}
\end{figure}

At the other end of the solid solution, het\ae rolite \zmo\,has fewer and
weaker magnetic peaks. While extensive work has been done on the magnetic 
ordering of \textit{cubic} spinels where the spins are confined purely on the 
B sublattice and are strongly geometrically 
frustrated,\,\cite{baltzer_exchange_1966, martinho_magnetic_2001, schiessl_magnetic_1996} 
the magnetic ordering in tetragonally 
distorted hausmannite/het\ae rolite B-site compounds has received less 
attention. There are three relevant tetragonal spinels to 
consider: \zmo, CdMn$_2$O$_4$, and MgMn$_2$O$_4$. Zn and Cd both have a strong 
tendency to occupy tetrahedral sites, but Mg is exhibits about 10\,\%-25\,\%
inversion on the octahedral sites.\,\cite{miller_distribution_1959}
No description of the magnetic structure has accompanied studies of
(Zn,Cd)$_x$Mn$_{3-x}$O$_4$.\,\cite{rosenberg_magnetic_1964,troyanchuk_phase_1996} 

To better understand the magnetic structures that are plausible with the
data, we display various depictions of the het\ae rolite crystal structure in
the panels of Fig.\,\ref{fig:structure}. The B-site octahedral cation 
sublattice displayed in Fig.\,\ref{fig:structure} can be described in two 
ways: as a pyrochlore lattice stretched in the $c$ direction, or as layers of 
parallel chains stacked at 90$^\circ$ to each other. In cubic spinels with 
nonmagnetic A-sites, the intrachain B--B direct exchange is the strongest 
magnetic interaction and is geometrically frustrated since it occurs within
ideal tetrahedra.\,\cite{ramirez_strongly_1994} In \zmo\,as in Mn$_3$O$_4$, 
the elongation along $c$ stretches two of the pyrochlore-type B-site nets and 
leaves one (in the $ab$ plane) unchanged. This has led to the interpretation 
of the het\ae rolite magnetic structure to consist of ferromagnetic chains of
Mn$^{3+}$, with antiferromagnetic interchain 
interactions.\,\cite{aiyama_magnetic_1966} This simple interpretation clearly 
does not capture all the details as is evident in the TOF neutron diffraction 
data, where the peaks in \zmo\,are diffuse and therefore indicate a 
substantial amount of disorder over long length scales. There is a shift of 
intensity from the (101) peak at $d$ = 4.9\,\AA\/ once $x$ increases past 0.5, 
and the intensity of the diffuse peak at $d$ = 5.05\,\AA\/ gradually increases 
until \zmo\,is reached. In the middle compound with $x$ = 0.5, no magnetic 
Bragg peakss are present.  There is only a slight increase in diffuse 
intensity around $d$ = 5\,\AA, so any magnetic order at this point must only 
be short-range in nature. 

\begin{figure}
\centering\includegraphics[width=8cm]{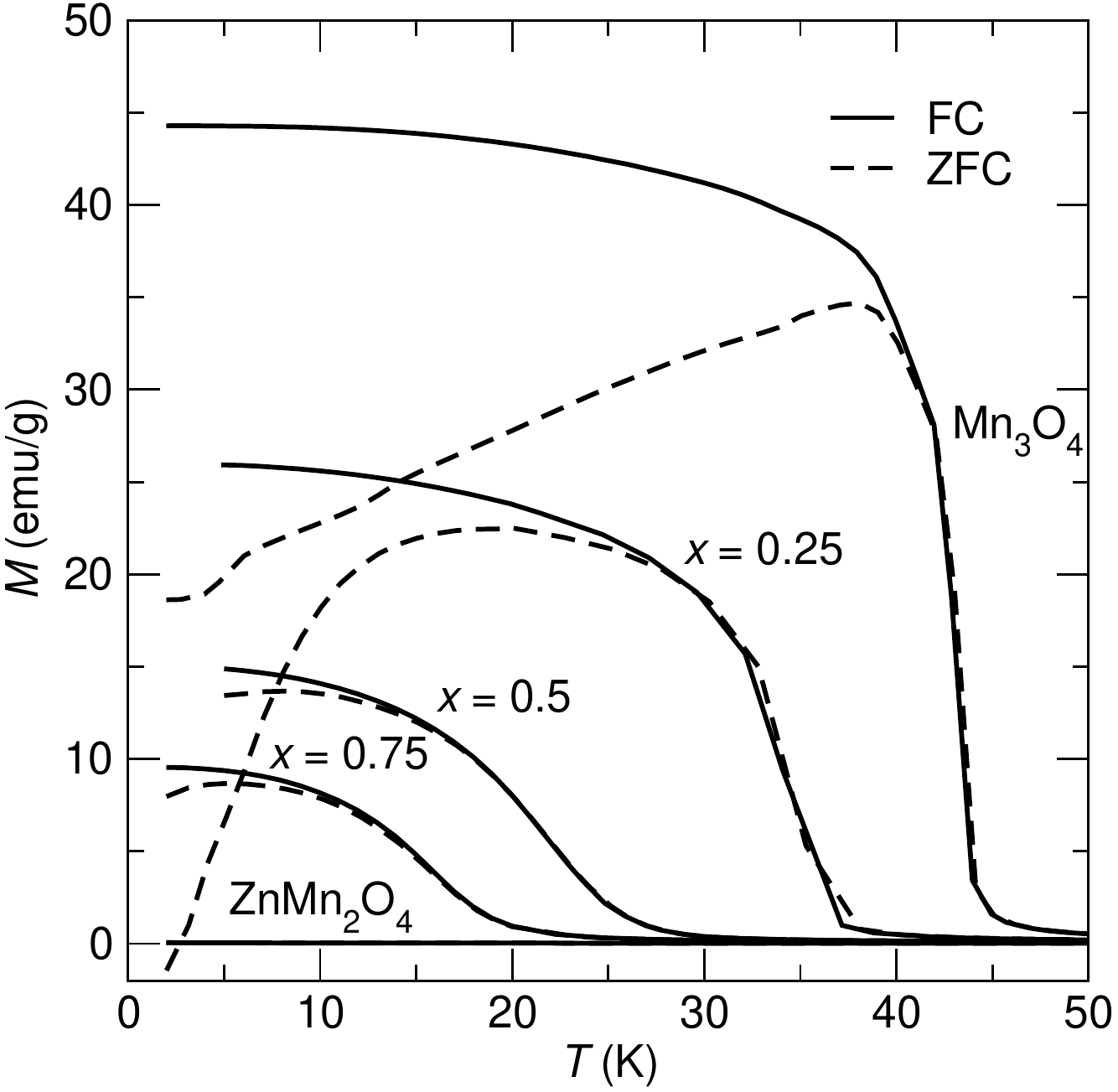}\\
\caption{Field-cooled (FC) and zero-field-cooled (ZFC) magnetization curves 
at $H$ = 1000\,Oe for the \zmox\,solid solution show a gradual decrease in 
the magnetic ordering temperature, as well as the magnetization 
from $x$ = 0 to 1. The interactions in ZnMn$_2$O$_4$ are 
antiferromagnetic changes cannot ben observed on this magnetization scale; 
these shown in greater detail in Fig.\,\ref{fig:zmo_zfc-fc}.} 
\label{fig:zfc-fc}
\end{figure}

While the magnetic neutron scattering data requires a more detailed
analysis that will be presented in future work, we use the general trends
to explain AC and DC magnetization measurements 
presented in this work. DC magnetization measurements on members of the
\zmox\,solid solution indicate a smooth, linear decrease in both the magnetic
ordering temperature as well as the maximum magnetization on going from 
Mn$_3$O$_4$ ($x$ = 0) to ZnMn$_2$O$_4$ ($x$ = 1).  The
field-cooled (FC) and zero-field-cooled (ZFC) magnetization curves in
Fig.\,\ref{fig:zfc-fc} show a steady decline in the ordering temperature,
temperature of magnetic irreversibility (deviation of ZFC and FC curves), and
FC moment as $x$ goes from 0 to 0.75.  The magnetization curves show that the
neutron TOF data in Fig.\,\ref{fig:hipd} at 20\,K is below $T_C$ for the four
ferrimagnetic samples.  The samples at $x$ = 0.5 and 0.75 have significant
diffuse intensity at 50\,K, well above $T_C$ measured \textit{via} SQUID
magnetization.  Interestingly, the weak magnetic scattering intensity in 
$x$ = 0.5 versus $x$ = 0.75 (Fig.\,\ref{fig:hipd}) seems contradict the fact 
that $x$ = 0.5 has the larger magnetization and higher $T_C$.  We can 
therefore assume that in $x$ = 0.5 samples, ferrimagnetism is caused by local 
regions of aligned spins which lack long-range order.  

\begin{figure}
\centering\includegraphics[width=8cm]{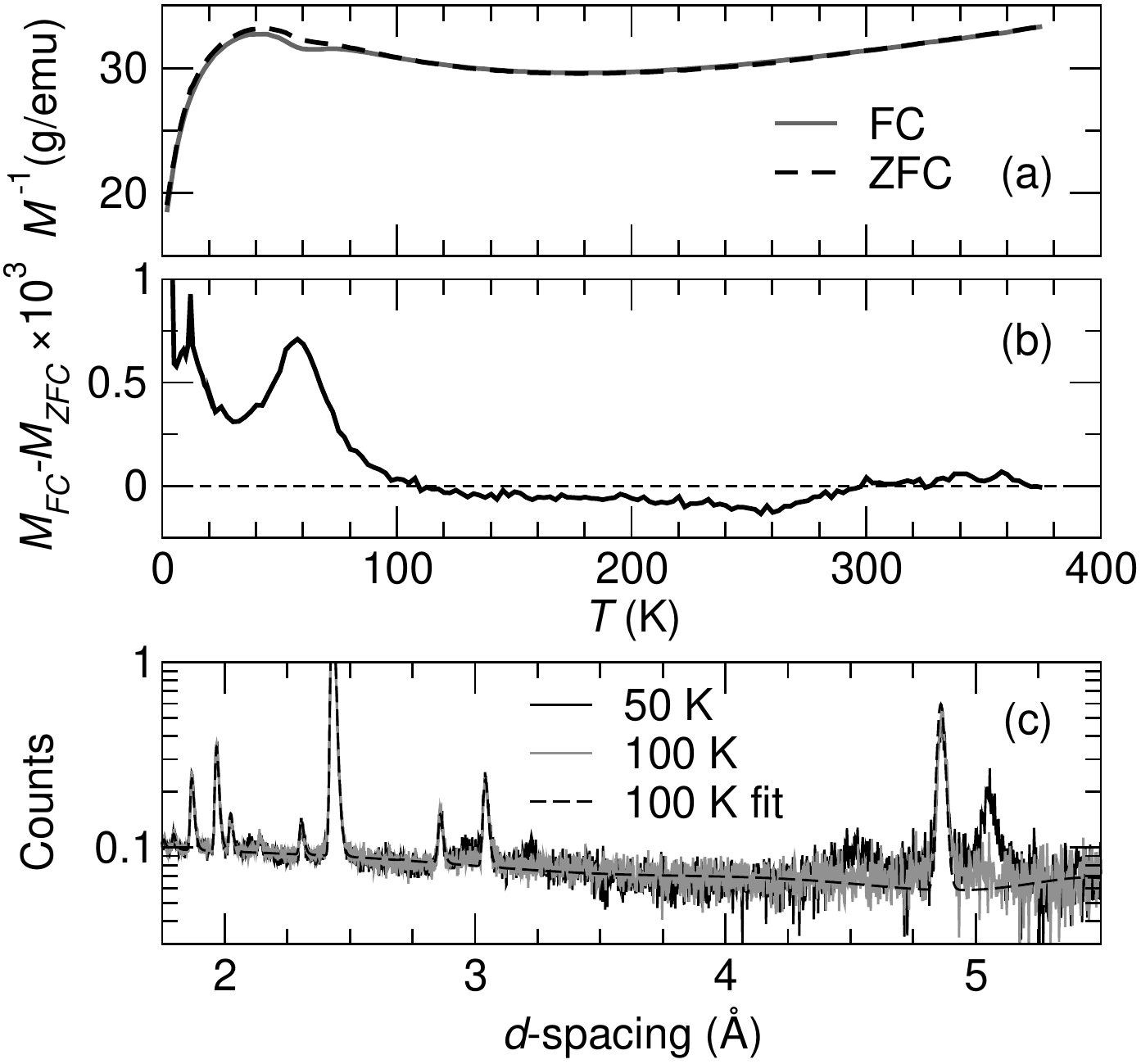}\\
\caption{(Color online) Inverse susceptibility ZFC/FC data (a) for \zmo\/ shows
Curie-Weiss behavior above room temperature with a very broad, gradual ordering
 of the spins that begins around 260\,K. Small amounts of irreversibility are 
seen in (b), which indicates a magnetic transition at $T$ = 60\,K. In (c), the 
appearance of a magnetic Bragg peak in TOF neutron data between 100 and 50\,K 
indicates the onset of long-range magnetic order coinciding with the peak in 
(b). The antiferromagnetic downturn in this sample only occurs at near 40\,K.
The Rietveld fit at 100\,K is for structural peaks only.}
\label{fig:zmo_zfc-fc}
\end{figure}

The ZFC-FC behavior for \zmo\,is much more complex than the other samples in 
the solid solution, and has been the subject of continued investigation for 
many years.\cite{jacobs_evidence_1959,rosenberg_magnetic_1964,aiyama_magnetic_1966,wautelet_interpretation_1974,wautelet_etude_1974,bhandage_magnetic_1978}
Salient features that have remained consistent are Curie-Weiss paramagnetism 
above room temperature, with a phase transition between 230\,K and 290\,K that 
has been detected in specific heat\cite{aiyama_magnetic_1966,chhor_heat_1986} 
and Young's modulus\cite{troyanchuk_phase_1996} measurements.
In our measurements of the ZFC/FC behavior in Fig.\,\ref{fig:zmo_zfc-fc}, we 
observe this as a gradual slope change in $M^{-1}$ versus $T$.
The irreversible moment $M_{FC} - M_{ZFC}$ has a slight dip around 260\,K and 
a strong transition at 60\,K. A new magnetic Bragg peak at $d = $ 5.05\,\AA\/
clearly arises between 100\,K and 50\,K and persists down to 20\,K.

\begin{figure}
\centering\includegraphics[width=8cm]{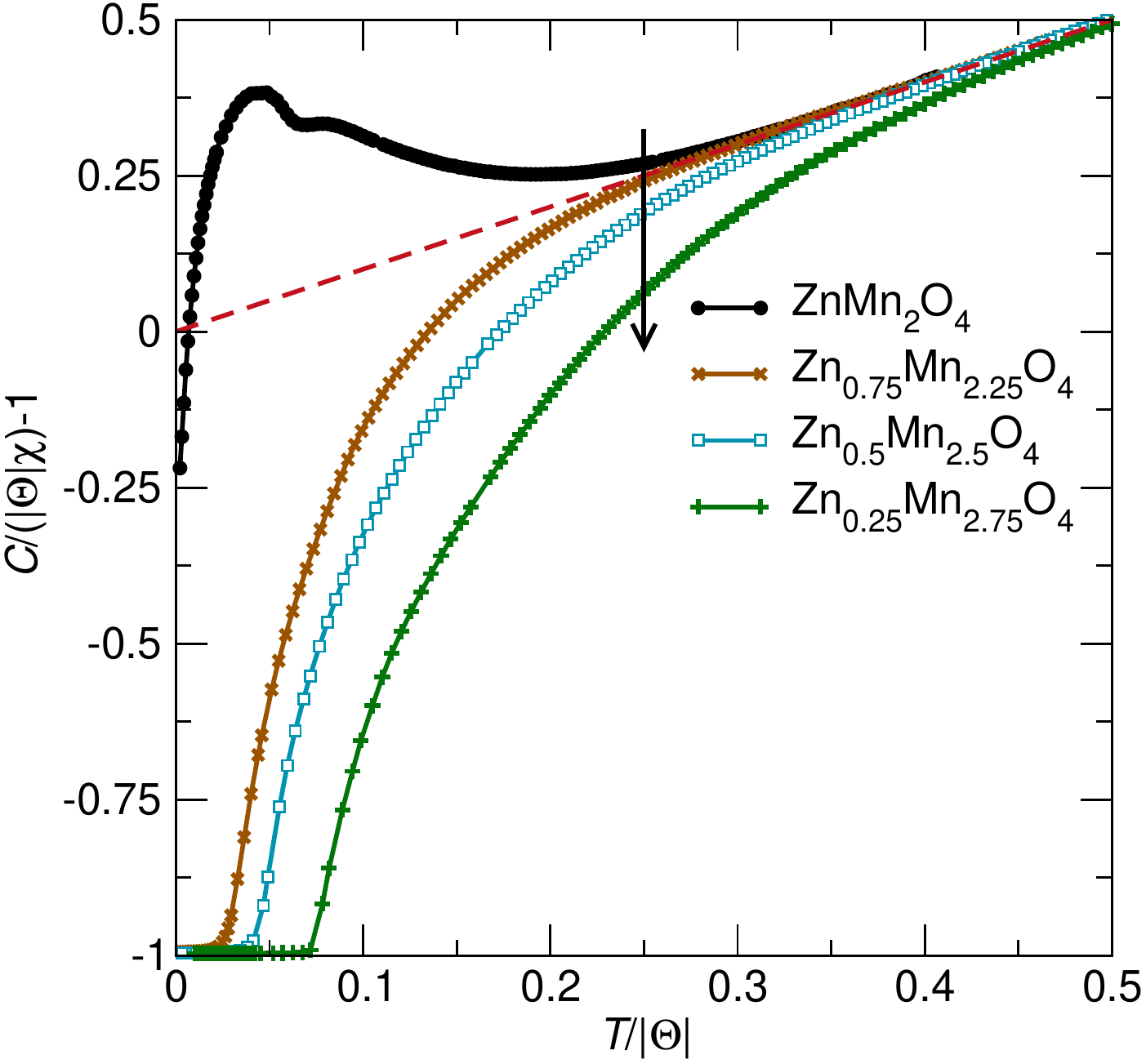}\\
\caption{(Color online) Curie-Weiss normalization of the FC magnetization 
curves provides a view of the differing magnetic ordering schemes in the 
\zmox\/ solid solution. Deviation from purely paramagnetic behavior (dashed) 
is ferrimagnetic for samples with $x < 1$, with $T_C$ decreasing with the 
number of A-site spins. Only \zmo\,has antiferromagnetic ordering at low 
temperature.} 
\label{fig:c-w}
\end{figure}

\begin{figure}
\centering\includegraphics[width=8cm]{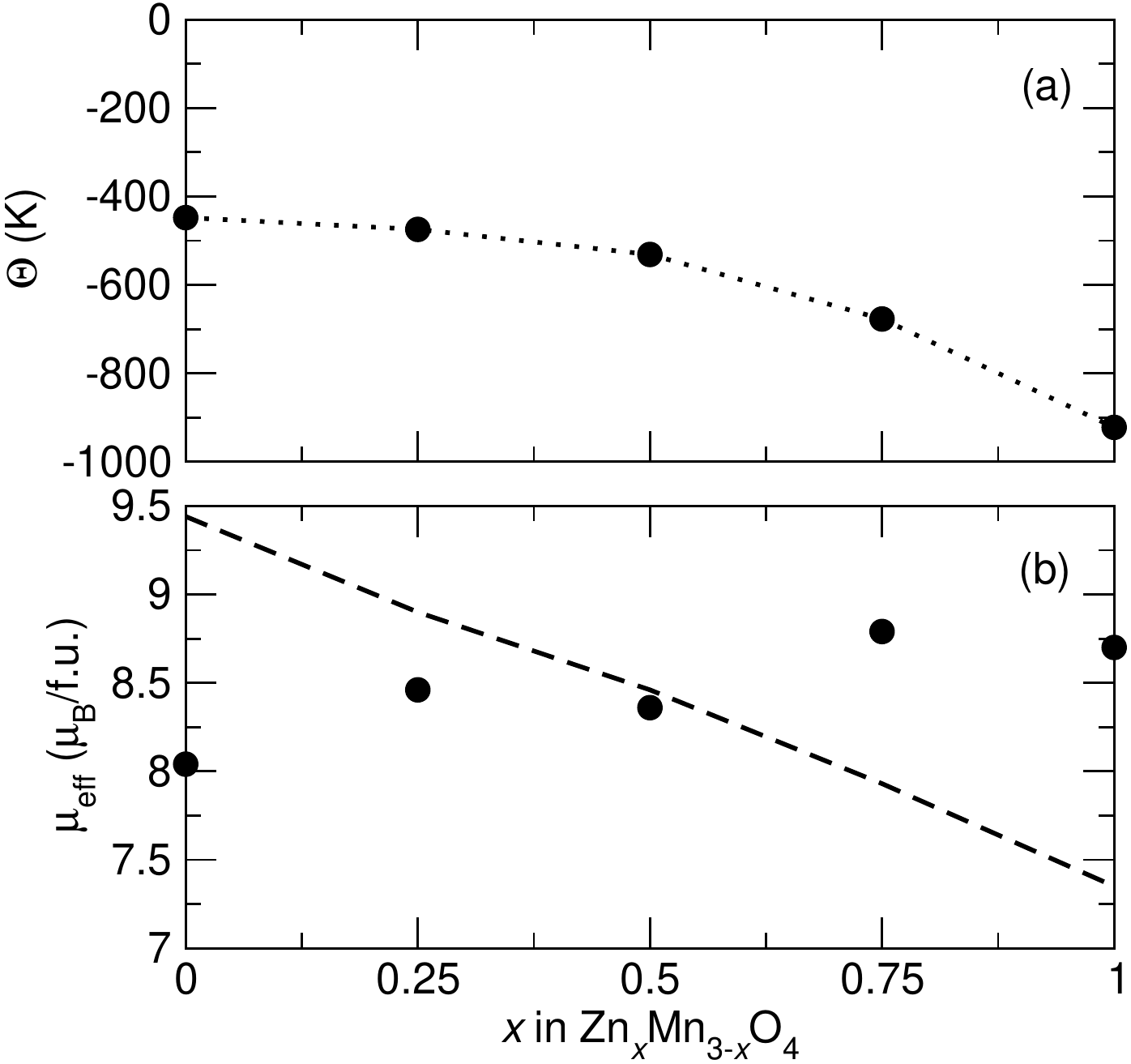}\\
\caption{The Curie-Weiss temperature $\Theta$ versus composition (a) shows 
increasing dominance of short-range antiferromagnetic interactions as the 
solid solution progresses from \haus\/ to \zmo. The dotted line is a guide to 
the eye. The paramagnetic $\mu_{eff}$ shown in (b) begins below the ideal 
$L+S$ contribution (dashed line) for \haus, but increases past the expected 
value for \zmo. This increase in effective moment with $x$ is counterintuitive 
since Mn$^{2+}$ spins are being \emph{removed}, but could be attributed to 
Jahn-Teller orbital ordering contributions.} 
\label{fig:theta-ueff}
\end{figure}

As Mn$^{2+}$ is substituted into the end member \zmo, ferrimagnetism is induced
and can be illustrated by normalizing the FC magnetization using the 
results of fitting the high-temperature susceptibility to the Curie-Weiss law.
The data are then displayed on a common scale, presented in Fig.\,\ref{fig:c-w}.
The utility of such scaling across solid solutions has proven crucial in 
previous studies of because it offers a clear view of relative strengths of 
FM/AFM interactions in similar compounds.\,\cite{melot_magnetic_2009}
All samples have Curie temperatures $\Theta <$ 0\,K, indicating that 
short-range  interactions are predominantly antiferromagnetic.
The trend of $\Theta$ versus $x$ is shown in Fig.\,\ref{fig:theta-ueff}(a).
The strength of antiferromagnetic coupling gradually increases as Zn is 
added to the A-sites, possibly as a consequence of the smaller cell volume
as Zn${2+}$ substitues Mn$^{2+}$. For \zmox\/ samples with $x < 1$, these 
interactions lead to ferrimagnetic order (dropping below the dashed line of 
ideal Curie-Weiss paramagnetism) with an ordering temperature that decreases 
with the concentration of tetrahedral Zn$^{2+}$. 

A more curious trend develops in the paramagnetic effective moment 
$\mu_{eff}$ which is measured above 300\,K. In Fig.\,\ref{fig:theta-ueff}(b), 
\haus\,has $\mu_{eff}$ = 8.04\,$\mu_B$/f.u. instead of the ideal value of 
9.44 for one tetrahedral Mn$^{2+}$ and two octahedral Mn$^{3+}$ per formula 
unit (including both spin and orbital contributions).  Interestingly, the 
experimental $\mu_{eff}$ increases with Zn content, despite the removal of 
$d^5$ Mn$^{2+}$.  If the discrepancy from the ideal value were due to 
short-range ordering in \zmo, we would expect \emph{lowering} of $\mu_{eff}$, 
but this is not the case.

\begin{figure}
\centering\includegraphics[width=8cm]{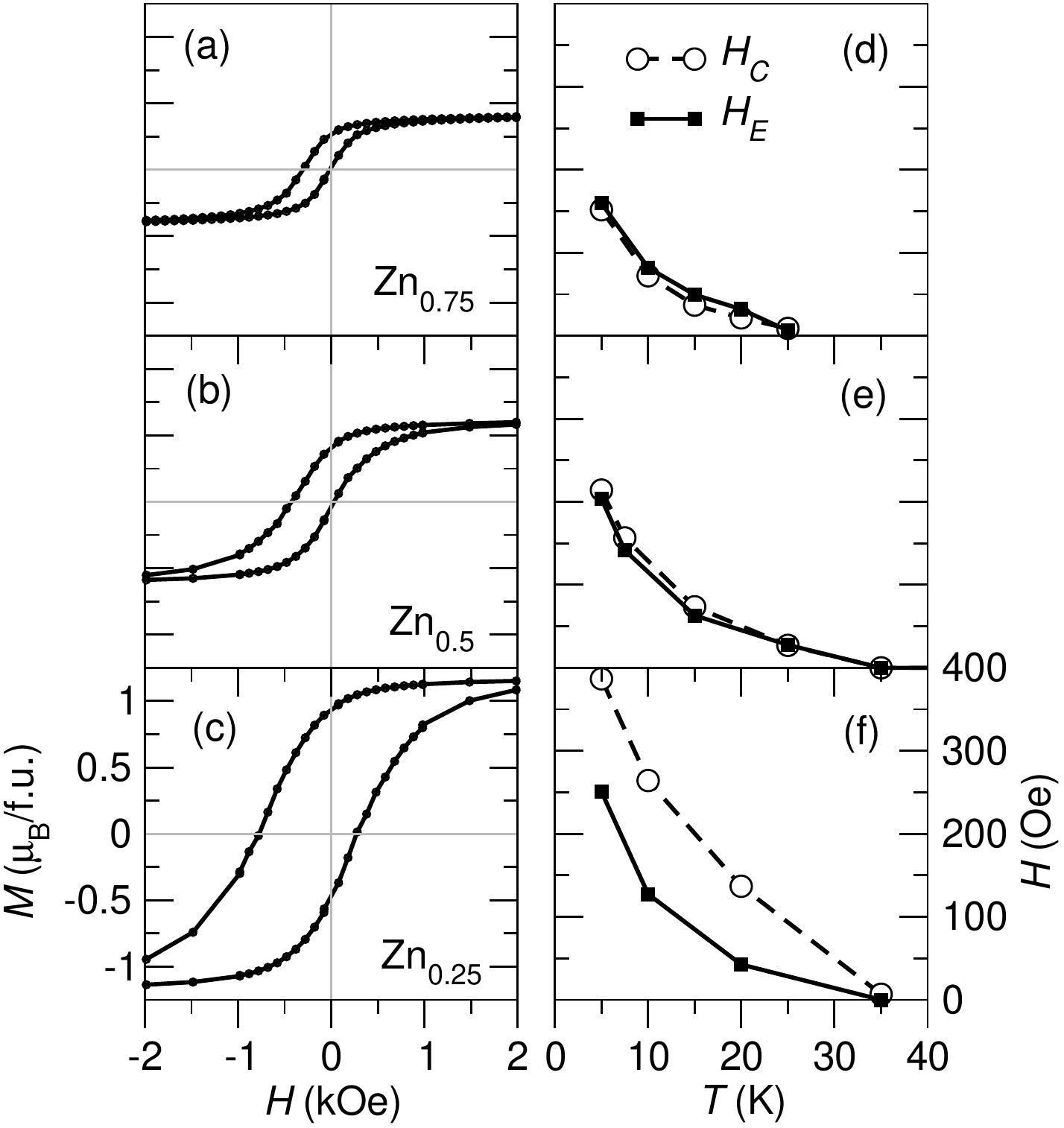}\\
\caption{Hysteresis loops (a-c) measured at 5\,K after $H_{FC}$ = +50\,kOe 
field-cooling show dramatic exchange-biased loop shifts. The $x$ = 0.75 and 
0.5 loops are pinned so that the coercive field $H_C$ in the $+H$ direction 
is zero. This results in overlapping values of loop shift $H_E$ and half loop 
width $H_C$ \textit{versus} temperature (d-f). Some shift is still evident in 
$x$ = 0.25 and disappears in Mn$_3$O$_4$.} 
\label{fig:he_hc}
\end{figure}

All hysteresis loops measured after ZFC in this system are symmetric around 
the origin. However, FC loops for $0<x<1$ measured under a cooling field 
$H_{FC}$ = 50\,kOe are shifted by an exchange bias field $-H_E$, as seen in 
Fig.\,\ref{fig:he_hc}.  Such loop shifts along $H$ after field cooling are 
similar to what was first reported by Jacobs and 
Kouvel.\cite{jacobs_exchange_1961} A systematic examination of the behavior 
from $0 \leq x \leq 1$ reveals an interesting trend.
\zmo\,is antiferromagnetic and displays no hysteresis.
As Mn$^{2+}$ is inserted on the tetrahedral sites, ferrimagnetism arises with a linearly increasing saturation magnetization.
In the $x$ = 0.25 and 0.5 samples, the loop shift is exactly equal to the 
coercivity--that is, $H_E = H_C$ if we define $H_C$ to be half the loop width.  
This implies that for a positive $H_{FC}$, nearly \textit{all} Mn spins that 
contribute to the ferrimagnetic behavior are pinned in the $+M$ direction when 
$H_{FC}$ is first relieved. As the hysteresis continues to negative saturation 
and $H$ is increased from $-$50 to 50\,kOe, there reaches a point where all 
the Mn ferrimagnetic spins are exactly compensating.
This occurs at $H=0$. The magnetic saturation $M_S$ varies smoothly from 
\zmo\/ to Mn$_3$O$_4$, with a contribution of about 0.30(4)\/$\mu_B/$ per 
Mn$^{2+}$, which has $S$ = 5/2 and could contribute a maximum of 5 $\mu_B$.
Because the ferrimagnetic end member Mn$_3$O$_4$ also obeys this relationship, 
we assume that the inserted Mn$^{2+}$ create nanoscale clusters of Mn$_3$O$_4$ 
that are the dominant source of the total magnetic moment.
These local FM clusters must be contained within an antiferromagnetic matrix 
because the exchange bias behavior is genuine, as indicated by the field-cooled
loop shifting and centered ZFC loops.

As the tetrahedral Mn$^{2+}$ fraction increases past 50\,\%, the loop shift 
changes from $H_E = H_C$ to $H_E = 0$ for the end member Mn$_3$O$_4$. When 
$x$ = 0.75, $H_E$ is still present but the positive $H_C$ value no longer 
resides at $H=0$ as it does for the completely shifted $x$ = 0.5 and 0.25 
cases. For a diamond-type lattice such as the A-sites in spinel or 
het\ae rolite, the site percolation threshold is 
43\,\%.\,\cite{marck_percolation_1997}
As percolation on the tetrahedral sublattice is achieved, loop shifting 
decreases while $H_C$ and $M_S$ vary gradually. So only the dilute spins near 
edges of clusters are pinned during field cooling, and the pinning is overcome 
when the clusters grow large or coalesce.  

\begin{figure}
\centering\includegraphics[width=8cm]{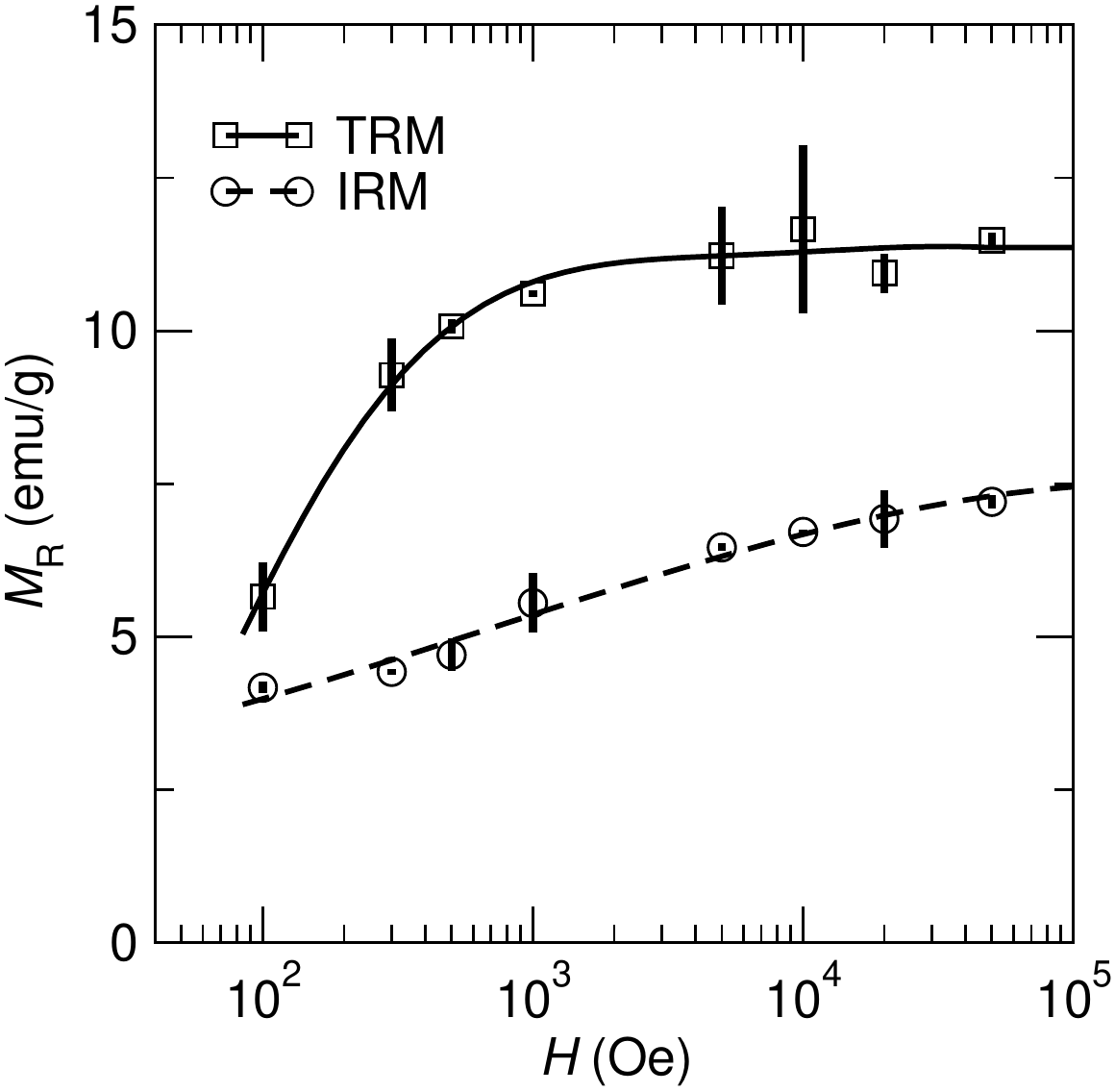}\\
\caption{Thermoremanent magnetization (TRM) and isothermal remanent 
magnetization (IRM) versus applied field for a $x$ = 0.5 sample shows clear 
deviation up to $H$ = 50\,kOe. Lines are guides to the eye. For a typical spin 
glass, the two curves should join with increasing $H$ as the field aligns the 
disordered moments to saturation. In an exchange biased system, the 
curves remain separated as seen here.} 
\label{fig:trm-irm}
\end{figure}

Loop shifts such as those in Fig.\,\ref{fig:he_hc} can arise from two 
phenomena: classical exchange biasing of a ferromagnet and antiferromagnet, or 
as a consequence of spin-glass behavior. In the latter case, $H_E$ can arise 
from coupling a ferromagnet to a spin glass,\cite{ali_exchange_2007} glassy 
uncompensated spins at interfaces/surfaces,\cite{martinez_low_1998} or an 
intrinsic anisotropy present in the glass 
itself.\cite{fischer_spin_1985,binder_spin_1986}
One method of testing for spin-glass behavior is the measurement of 
thermoremanent and isothermal remanent magnetization (TRM and IRM, 
respectively) shown in Fig.\,\ref{fig:trm-irm}.
The TRM measurement begins as a typical FC procedure: $H_{FC}$ is applied 
while cooling from above the magnetic transition, temperature is stabilized, 
$H_{FC}$ is removed, and the remanent moment $M_R$ is measured.
For an IRM measurement, the sample is zero-field cooled, the temperature is 
stabilized, $H$ is applied for a substantial length of time (here we use 
30\,min.), the applied field is removed, and $M_R$ is measured.
In glassy systems, TRM is greater than IRM for low $H_{FC}$ because additional 
alignment is induced while cooling through the high-susceptibility glass 
transition.\cite{tholence_susceptibility_1974,aharoni_isothermal_1984} 
At high $H_{FC}$ the values coincide when the applied field overcomes 
intrinsic anisotropy and aligns all spins, regardless of thermal history.
In an exchange biased material, antiferromagnetic spins are \textit{not} 
reversed by high fields, so the TRM and IRM curves remain separated even at 
high fields. Indeed, we can see in Fig.\,\ref{fig:trm-irm} that for \zmofiv\/
high values of $H_{FC}$ produce a higher value for the exchange-biased TRM 
than the ZFC, non-biased IRM. The TRM/IRM data disallows considering the 
A-site Mn$^{2+}$ spins to be a dilute ferromagnetic spin glass that are 
coupled to an antiferromagnetic B-site sublattice.
This measurement further corroborates a two-phase interaction between 
ferrimagnetic Mn$_3$O$_4$-type clusters with \zmo-type antiferromagnetic 
regions. 

Note that these phases are not ordered on the long range, as 
evidenced most clearly by the diffraction pattern for \zmofiv\/ in 
Fig.\,\ref{fig:hipd}(c).  The magnetic Bragg peaks disappear when $x$ = 0.5, 
even though the trends in SQUID magnetism continue to vary smoothly.
Nevertheless, the ferrimagnetism and exchange bias act as direct interpolations
of the $x$ = 0.25 and 0.75 samples. In \zmo\/ some magnetic ordering produces 
Bragg peaks, but a loss of Bragg intensity with $x$ signals the breakdown of 
this B-site ordering from the stronger (but still antiferromagnetic) A-B 
coupling to the inserted A-site Mn$^{2+}$. 

\begin{figure}
\centering\includegraphics[width=8cm]{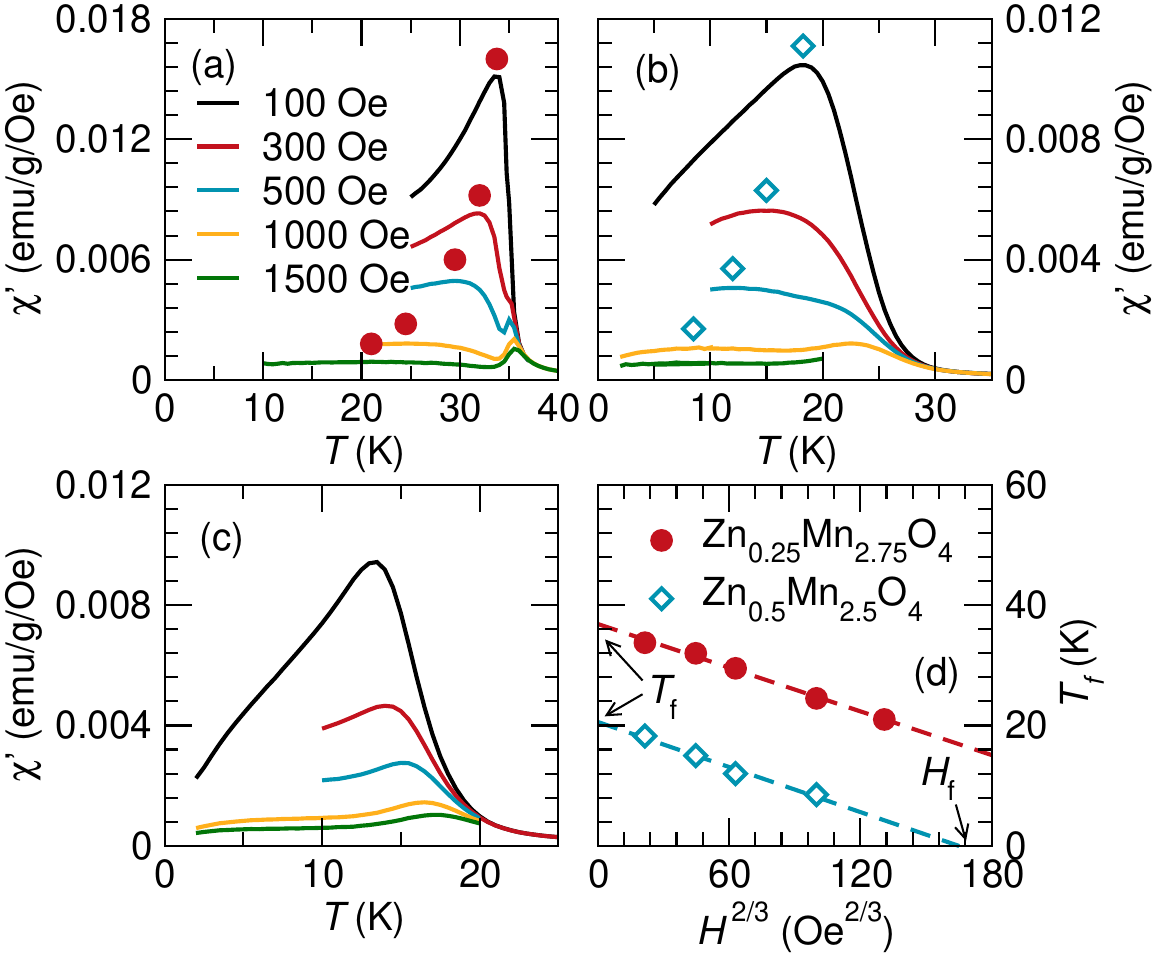}\\
\caption{(Color online) Magnetic AC susceptibility for with mixed tetrahedral 
occupancy: (a) \zmotwo, (b) \zmofiv, and (c) \zmosev. The AC field is 1\,Oe 
with different DC fields shown. Local maxima in (a) and (b) are marked with 
symbols and replotted in (d) to show de Almeida--Thouless behavior. No such 
trend is present in (c), where maxima are present only at the ferrimagnetic 
$T_C$ around 18\,K. Spin-glass freezing temperatures $T_f$ and critical fields 
$H_{cr}$ can be extracted for both curves in (d): for \zmotwo\,$T_f$ = 36.9 K 
and $H_{cr}$ = 5320 Oe; for \zmofiv\,$T_f$ = 20.6 K and $H_{cr}$ = 2020 Oe.}
\label{fig:ac}
\end{figure}

In the AC magnetization measurements of Fig.\,\ref{fig:ac}, two maxima are 
seen in $\chi^\prime$ under cooling: one at $T_C$ and another at a lower 
temperature, which is interpreted as a spin-glass freezing 
$T_f$.\cite{fischer_spin_1985, binder_spin_1986}
The glassy spins may be present at the interfaces between the A-site-induced 
ferrimagnetic clusters or (less likely) as isolated sites.
For samples with $x$ = 0.75 and 0.5 (Figs.\,\ref{fig:ac}a and \ref{fig:ac}b), 
$T_f$ shifts to lower temperatures as the DC bias magnetic field is increased.
The $T_f$ versus $H^{2/3}$ dependence plotted in Fig.\,\ref{fig:ac} indicates 
excellent agreement with de Almeida--Thouless (AT) 
behavior,\cite{almeida_stability_1978} which is typical not only for bulk 
frustrated and dilute spin glasses,\cite{efimova_frustrated_2005} but also for 
a wide variety of systems with disordered spins at surfaces and 
interfaces.\cite{martinez_low_1998,makhlouf_magnetic_1997,gruyters_spin_2005}
No such behavior is seen in the $x$ = 0.25 sample, since the Mn spins now 
occupy 75\,\% of the A-sites and the ferrimagnetic phase has effectively 
percolated the entire structure.
Two key values can be extracted from the AT lines in Fig.\,\ref{fig:ac}(d): 
the freezing temperature $T_f$ where irreversibility (hysteresis) in the spin 
glass is first induced, and the critical field $H_{cr}$ where the applied 
field overcomes the internal anisotropy of the spin glass and saturates it.
Considering \zmofiv, $T_f$ = 20.6\,K, which is slightly higher than the DC 
deviation of ZFC-FC data in Fig.\,\ref{fig:zfc-fc}, as expected since the DC 
data was collected at $H$ = 1000\,Oe.
More importantly, $H_{cr}$ = 2020\,Oe. This implies that if the $M_R$ were 
solely due to a spin glass component the TRM-IRM curves would coincide at 
$H_{cr}$. As they do not, the number of glassy spins must be very small in 
comparison to those in ferrimagnetic clusters.
Thus the irreversible magnetization in the hysteresis loops of 
Fig.\,\ref{fig:he_hc}(b) primarily arises from ferrimagnetic regions of local 
spin alignment and not from glassy clusters that obey AT behavior.

\begin{figure}
\centering\includegraphics[width=8cm]{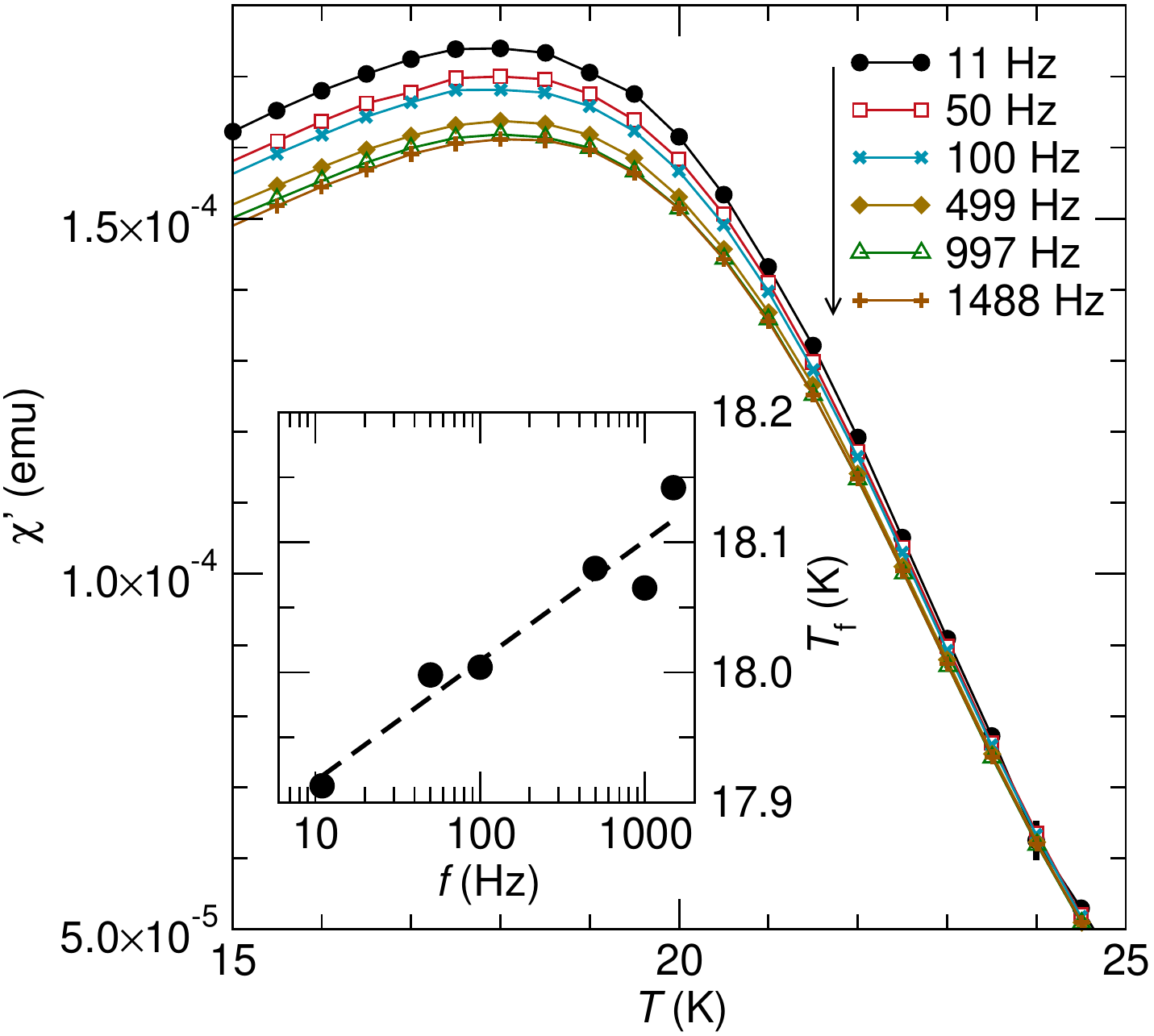}\\
\caption{(Color online) The AC magnetic susceptibility for \zmofiv\/ exhibits 
frequency dependence in the region associated with spin glass freezing.
The $T$-value of the maximum is plotted versus $f$ in the inset.
Error bars are smaller than the data points. The variation of $T_f$ with $f$ 
agrees with standard spin glass behavior. The $T_g$ extracted from this data 
differs from that in Fig.\,\ref{fig:ac} due to the large non-glassy 
ferrimagnetic contribution.} 
\label{fig:ac-freq}
\end{figure}

Frequency-dependent AC magnetization measurements of the $T_f$ region in 
\zmofiv\/ in Fig.\,\ref{fig:ac-freq} show a deviation after cooling below 
$T_f$, further evidence of a small amount of glassy behavior. The peak centers
are plotted versus 
$f$ in the inset. The cusp in $\chi^\prime$ obeys the relationship 
$\Delta T_f/[T_f(\log \omega)]$ = 0.005, which is the same value as the 
canonical spin glass CuMn.\,\cite{mydosh_spin_1993}
The breadth of the peak indicates that there is a distribution of freezing 
temperatures, based on the non-uniform distribution of glassy spins on 
interfaces of the ferrimagnetic clusters.

\section{Conclusions} 

The system \zmox\/ is a homogeneous solid solution when investigated using
bulk structural probes such as TOF neutron diffraction. However, magnetic 
measurements reveal intrinsic exchange bias that we believe results from the 
interaction of distinct ferrimagnetic and antiferromagnetic regions. For 
concentrations of Mn-doping up to 50\,\%, field-cooled hysteresis loops are 
shifted so that $H_E = H_C$. Because magnetic scattering is diffuse, and 
the Curie-Weiss temperature $\Theta$ is large and negative, the magnetic 
structure of the \zmox\/ solid solution must consist of ferrimagnetic Mn-rich 
clusters that do not order on a macroscopic scale. As the clusters grow, their 
contribution to $M_S$ increases linearly until \haus\,is reached, and exchange 
bias disappears. There is a glassy component to the the magnetism in these 
systems, as evidenced by AC magnetization measurements.  
However, the contribution of glassy spins to the DC magnetization is minimal, 
which is most visible in the well-separated TRM and IRM traces even up to 
large fields. The presence of intrinsic exchange bias merits further 
investigation of the nanoscale ordering of spins in the \zmox\/ system.
Small-angle neutron scattering, real-space total scattering, 
Lorentz transmission electron microscopy, and magnetic force microscopy could 
each help observe the evolution of magnetic ordering as a function of 
temperature and composition in this solid solution.

\section{Acknowledgments}
We thank B. C. Melot for helpful discussions. This work was supported by the 
Institute for Multiscale Materials Studies, the donors of the American Chemical
Society Petroleum Research Fund, and the National Science Foundation through
a Career Award (DMR 0449354) to RS and for the use of MRSEC facilities 
(DMR 0520415). Neutron scattering was performed on HIPD at the Lujan Center at 
the Los Alamos Neutron Science Center, funded by the DOE Office of Basic 
Energy Sciences. Los Alamos National Laboratory is operated by Los Alamos 
National Security, LLC under DOE Contract DE-AC52-06NA25396.

\bibliography{zmo}

\end{document}